\documentclass[aps,prl,twocolumn]{revtex4}
\usepackage{xcolor}
\usepackage{bm, graphicx, amsmath}
\usepackage{bbm}
\usepackage{amssymb,amsmath}
\usepackage[mathscr]{eucal}
\usepackage{graphicx}
\renewcommand{\vec}[1]{\mbox{\boldmath$#1$}}

\newcommand{\tr}{{\rm tr}\,}
\newcommand{\ket}[1]{\left|{#1}\right\rangle}
\newcommand{\bra}[1]{\left\langle{#1}\right|}
\newcommand{\braket}[2]{\langle{#1}|{#2}\rangle}

\DeclareMathOperator*{\argmax}{arg max}

\newtheorem{theo}{Theorem}

\newtheorem{prop}{Proposition}
\begin{document}

\title
{The quantum change point}
\author{Gael Sent\'{\i}s$^1$, Emilio Bagan$^2$, John Calsamiglia$^2$, Giulio Chiribella$^{3,4}$, and  Ramon Mu\~{n}oz-Tapia$^{2}$}
\affiliation{
$^{1}$Departamento de F\'{i}sica Te\'{o}rica e Historia de la Ciencia, Universidad del Pa\'{i}s Vasco UPV/EHU, E-48080 Bilbao, Spain\\
$^{2}$F\'{i}sica Te\`{o}rica: Informaci\'{o} i Fen\`{o}mens Qu\`antics, Departament de F\'{\i}sica, Universitat Aut\`{o}noma de Barcelona, 08193 Bellaterra (Barcelona), Spain\\
$^{3}$Department of Computer Science, The University of Hong Kong, Pokfulam Road, Hong Kong\\
$^4$ Canadian Institute for Advanced Research,
CIFAR Program in Quantum Information Science, Toronto, ON M5G 1Z8}

\begin{abstract}
Sudden changes are ubiquitous in nature.  Identifying them   
is   crucial  for a number of applications   in   biology, medicine, and social sciences.  Here we take    the  problem of detecting sudden changes to the quantum domain. We consider a source that emits quantum particles in a default state, until a
point where a mutation occurs that causes the source to switch to another state.
The problem is then to find out where the change occurred. 
    We determine the maximum  probability of correctly  identifying the change point, allowing for    
  collective measurements on the whole sequence of particles emitted by the source.   Then, we devise online strategies
    where  the particles are measured individually and  an  answer is provided as soon as  a new  particle is received.   We show that these online strategies substantially underperform  the optimal  quantum measurement, indicating  that quantum sudden changes, although happening locally, are better detected globally.  

\end{abstract}


\maketitle

The detection of sudden changes in a sequence of random variables is a pivotal topic in statistics,  known as the \emph{change point problem}~\cite{page, brodsky, nikiforov}.    The problem has widespread applications, including the study of  stock market variations~\cite{chen-2},  protein folding~\cite{pirchi}, and  landscape changes~\cite{ficetola}.  In general, identifying  change points plays a crucial role in all problems involving  
the  analysis of samples collected over time~\cite{brodsky,chen} because such analysis requires the stability of the system 
parameters~\cite{hawkins}. If changes are correctly detected, the sample can 
be conveniently divided in  subsamples, which can then be analyzed by the standard statistical techniques. 
 The detection of change points can also  be viewed as a border problem~\cite{rosten}, namely a problem where one wants to draw a   separation between two (or more) different configurations ---a task that  plays a central role in machine learning~\cite{takeuchi}.

\begin{figure}  
\begin{center}
  \includegraphics[width=1\linewidth]{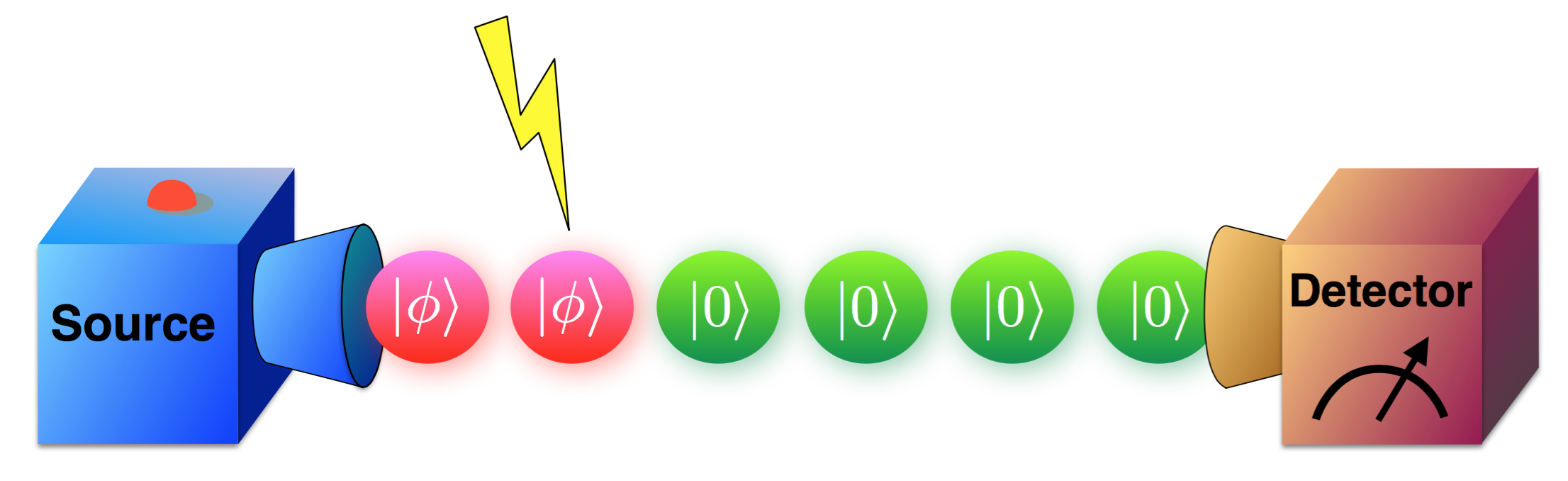}
  \end{center}
\caption{\label{fig:changepoint}
  {\bf The quantum change point problem. }   A quantum source emits particles in a default state $|0\rangle$, until the point where   a mutation  occurs, causing the source to emit particles in a different state $|\phi\rangle$.  A detector receives the stream of particles emitted from the source and measures them, producing an estimate of   the point where  the mutation occurred. 
  }
\end{figure}

The simplest   example of a change point problem is  that of a coin with variable bias.
    Imagine that a  game of Heads or  Tails is played with a fair coin, but after a few rounds one  player suspects that the other has replaced the coin with a biased one.  After inspection of the coin, the suspicion is confirmed: the coin has now a bias.      Can we identify  when the coin  was changed based only on   the  the sequence of outcomes?   This classical problem has a natural  extension  to the quantum realm, illustrated in Figure \ref{fig:changepoint}:    A   source is promised  to prepare quantum particles in  some default state. At  some point,   however, the source undergoes a  mutation  and starts 
   to  produce  copies of a different state.  Given the sequence of particles emitted  by the source,  the problem is to find out when the change took place.    
In the  basic version of the problem, the initial and final states are  known, as in the classical example of the  coin. No  prior information is given  about the location of the change: a priori, every point of the sequence   is equally likely to be   the change point.    
For simplicity, we assume    the quantum states  to  be pure.    More elaborate variations can be considered, e.g.,~with unknown states, non-uniform priors, mixed states, and multiple change points.   However, as we will see in this Letter,  the basic scenario already captures  the  intriguing  features that distinguish the quantum change point problem from its classical version.     An example   where the change can happen only at two possible positions and  the states are completely unknown was studied in ~\cite{akimoto},
 where the problem was shown to be equivalent to programmable discrimination~\cite{gael-1}.   

The quantum change point problem can be formulated as  a problem of state discrimination.  For a sequence of $n$ particles emitted by the source, the problem of  is to distinguish among $n$ quantum states, the $k$-th state having the change point in the $k$-th position.  Notably, state discrimination problems  with multiple quantum states have no closed-form solution  in  general    (see \cite{nakahira} for  some recent progress).  A complete solution is known only for the two-state case,  a result  that dates back to Helstrom's seminal work four decades ago \cite{helstrom} ---and even in this case computing the success probability may not be straightforward, see, e.g., the derivation of the quantum Chernoff bound \cite{chernoff}. 
For $n\ge 3$, the only cases where a solution is known are those of pure states with a high degree of symmetry. This includes  the symmetric states \cite{symmetric}, generated by the action of a unitary operation $U$ satisfying $U^n= \openone$,  and,  more generally,  states that are generated by a group of unitaries  \cite{chi2004,chi2005}.  

Interestingly, the change point problem does not fall into any of the above categories.  In spite of this, we show that the problem can be completely solved in the asymptotic regime: in the limit of long sequences,   the maximum probability of success  
 takes the elegant form 
\begin{align}\label{uno}
P_{\max}  =\frac{4(1-c^2)}{\pi^2}  \, K^2\left(c^2\right) +  O\left(\frac 1{n^{1-\epsilon}}\right) \, ,
\end{align}
where $c$ is the overlap between the default state and the alternate state,    $K(x)$ is the complete elliptic function of the first kind~\cite{abramowitz}, and $\epsilon>  0$ is an arbitrary  constant.  Quite remarkably,  the limit probability has a  non-zero value despite the fact that the number of states to be distinguished tends to infinity.   

Eq.~(\ref{uno}) characterizes the ultimate quantum limit to the detection of a change point.   Achieving the limit requires a nonlocal measurement, performed jointly on all the particles in the sequence.   To perform  such measurement one needs a quantum memory, wherein the states received from the source can be stored as they arrive to the detector.   Since quantum memories are challenging to implement,   it is  also interesting to consider local strategies,  where the particles are measured as soon as they arrive to the detector,  possibly adapting the measurement settings at one step based on the  outcomes of the previous measurements steps. 
Local strategies are interesting also because they can  provide an online answer:  they have a chance to identify the change point as soon as it occurs, without having to  wait until all the $n$ particles are scanned. 
When the online strategies are compared with the optimal quantum strategy, our results indicate the presence of a gap, showing  that the availability of a quantum memory and  the ability to perform nonlocal measurements offer  an advantage in the identification of the change point.



Let us  derive the optimal quantum strategy.  We denote by $\ket{0}$ the default state  and 
 by $\ket{\phi}=c \ket{0}+\ s \ket{1}$ 
 the state after the change. Without loss of generality, we choose $c$ to be real and positive.   
   If  the change occurs in the  position $k$,  the state of the $n$ particles is 
\begin{equation}
\label{psik}
\ket{\Psi_k}=   |0\rangle^{\otimes {k-1}}    |\phi\rangle^{\otimes n-k +1}
\end{equation}
%
We call the above states the \emph{source states}.   Note that the source states are linearly independent, except in the trivial case  when the states $|\phi\rangle$ and $ |0\rangle$ are equal.   In principle, the change can  occur  in any position, meaning that  every source state   has the same \emph{a priori} probability $1/n$    \footnote{It is easy to include the case where the change does not occur at all, by adding the vector $|\Psi_0\rangle=   |0\rangle^{\otimes n}$ and replacing the probability with $1/(n+1)$.}.   
The detector is described by  a  positive operator-valued measure (POVM), namely  a set of operators $\{M_k\}_{k=1}^{n}$   satisfying the positivity condition $M_k\geq 0$ and the completeness relation $\sum_{k=1}^{n}M_k =\openone$, where   $\openone$ denotes the identity  in the space $\mathcal{S}$ spanned by the source states. 
The average probability of successfully identifying the change point is
$
P=(1/n)\sum_{k=1}^{n} \bra{\Psi_k} M_k \ket{\Psi_k}
$
and our goal is to maximize it over all possible POVMs.

As mentioned above, the source states do not fall into any of the categories of states that admit a closed-form solution to the detection problem.  Still, we now show how an optimal solution can be constructed  in  the large $n$ limit. The key to our argument is a general result about the discrimination of linearly independent pure states, which  is of independent interest: 
  
\begin{theo}\label{theo}
Let $\{  |\Psi_k\rangle\}_{k=1}^n$ be a set of linearly independent states and let 
\begin{align}G_{ij}   =  \langle \Psi_i|  \Psi_j\rangle
\end{align} 
be the components of the corresponding Gram matrix.     The maximum probability of correctly identifying a state drawn uniformly at random from the set  $\{  |\Psi_k\rangle\}_{k=1}^n$ satisfies the bounds 
\begin{align}\label{lower}
 P_{\max} \ge    \left(  \frac {\tr \sqrt G}n\right)^2 
 \end{align}
and 
\begin{align}\label{upper} 
P_{\max} \le  \left(  \frac {\tr \sqrt G}n\right)^2     +     \sqrt{\lambda_{\max }}    \,  \| {\bf q}  -  {\bf u} \|_1  \, , 
\end{align}  
where   $\lambda_{\max}$ is the maximum eigenvalue of $G$,     ${\bf q}   =  \{q_k\}$ is the probability distribution defined by  $q_k  :  =   \big(  \sqrt G\,\big)_{kk}/\tr[\sqrt G]$, $  {\bf  u}  =  \{u_k\}$ is the uniform distribution  ($u_k=  1/n$ for all $k$),  and $\|  {\bf q}  - {\bf u}\|_1  : =\sum_k  |q_k-u_k|$ is the trace norm. 
\end{theo}
The  proof of Theorem \ref{theo} is provided in the Supplemental Material~\cite{appendix}, where we further extend the result to  non-uniform  prior distributions.  
   Note that the two bounds~(\ref{lower}) and~(\ref{upper}) are exactly equal when  the diagonal matrix elements of $\sqrt G$ are all equal to each other. In this case, Theorem \ref{theo}    yields the exact value of the success probability, reproducing a recent result of Ref. \cite{pozza}. 
   
We  now evaluate the bounds (\ref{lower}) and (\ref{upper}) for the change point problem, showing that the two bounds match at the leading order.  We start by evaluating the trace of~$\sqrt G$.    First, we observe that the Gram matrix has matrix elements  $G_{ij}  =   c^{|i-j|}$ and that   its inverse  has the simple form 
\begin{align}
\label{g-inv}
G^{-1}=\frac{1+c^2}{1-c^2}\,  \openone -\frac{c}{1-c^2} \,  H \, , 
\end{align}
where
$H_{ij}=\delta_{i\, j+1}+ \delta_{j\, i+1} + c(\delta_{i\, 1}\delta_{j\, 1}+ \delta_{i\, n}\delta_{j\, n} )$.
Luckily, the eigenvalues and eigenvectors of $H$ can be constructed explicitly:    
 in the Appendix we show that the eigenvalues have the form  $2\cos\theta_l$, where~$\theta_l$ is a suitable angle in the interval of size $\pi/n$ centered around~$\pi l/n$~\cite{appendix}. Eq. (\ref{g-inv}) then implies that the eigenvalues of the Gram matrix $G$ are
\begin{equation}
\label{lambda}
\lambda_l=\frac{1-c^2}{1-2\,c\,\cos\theta_l +c^2} \, ,   
\end{equation}
so that  we have 
\begin{align}
\label{ps-discrete}
\frac{  \tr \sqrt G}  n  =  
\frac{1}{n}\sum_{l=1}^n \sqrt{\frac{1-c^2}{1-2\,c\,\cos\theta_l +c^2}}\, .
\end{align}
Since the angles $\theta_l$ are distributed in intervals of equal size, forming a partition of  the interval $[0,\pi)$,  
 the sum can be replaced by  an integral  in the large $n$ limit, yielding the asymptotic equality
 \begin{align}
\nonumber 
\frac{\tr \sqrt G}n   &=   
\frac{\sqrt{1-c^2}}{\pi}\int_{0}^\pi \frac{d\theta}{\sqrt{1-2\,c\,\cos\theta +c^2}} \\
     \label{ps-cont}   &=\frac{2\sqrt{1-c^2} }{\pi} K\left(c^2\right) \, ,
\end{align}
valid up to an error of size $1/n^{1-\epsilon}$ \cite{appendix}. 
According to Eq. (\ref{lower}), the square of the r.h.s. is a lower bound for the maximum success probability.
   
%

Let us now evaluate the upper bound (\ref{upper}).  First, note that we have  $\lambda_{\max}\le (1+c)/(1-c)$, as one can easily  read out from Eq. (\ref{lambda}).   Moreover, it is possible to show that the probability distribution $\bf q$ is approximately uniform, with the bound  $\|  {\bf q}  - {\bf u}\|_1  \le   {4  (1+c)}/{ (1-c)} \,    1/{n^{1-\epsilon}}$ holding at the leading order in $n$ \cite{appendix}.   Hence, the upper bound (\ref{upper})
 yields     the   inequality
 \begin{align}P_{\max} \le \left ( {\tr\sqrt G}/n  \right)^2    +  4  \left(  \frac{ 1+c}{1-c}\right)^{3/2}    \frac 1{n^{1-\epsilon}}  \, .
 \end{align}   
 In summary, the bounds (\ref{lower}) and (\ref{upper}) match in the asymptotic limit, up to an error of size $1/n^{1-\epsilon}$. This establishes the validity of Eq. (\ref{uno}).

\begin{figure}
   \centering
   \includegraphics[width=3.3 in]{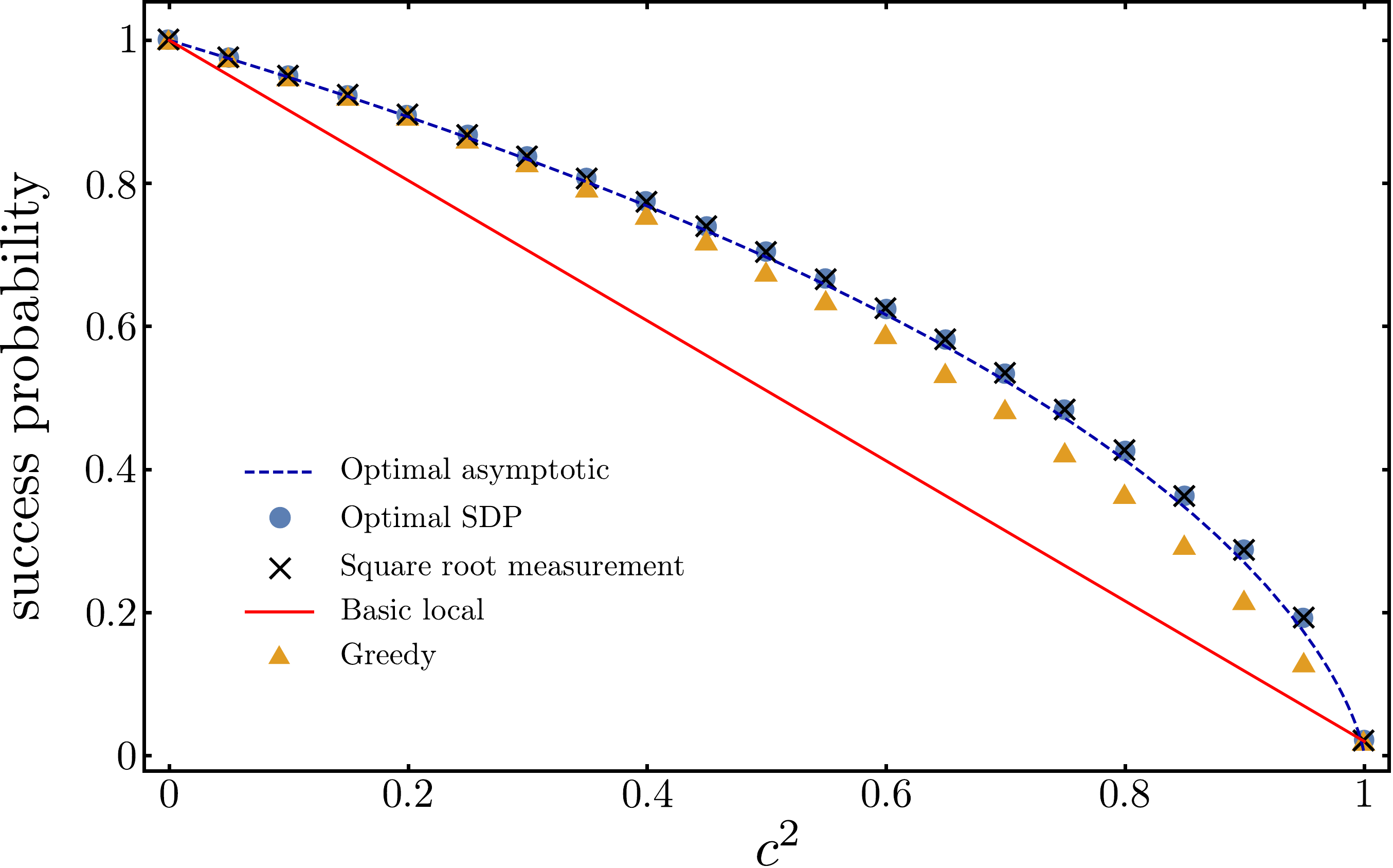} 
\caption {Probability of correct identification of the change point as a function of  $c^2=|\braket{0}\phi|^2$.  The (blue) dashed line is the asymptotic result in 
Eq.~\eqref{uno}. For a sequence of $n = 50$ states, we also plot the maximum probability, obtained by SDP optimization, (blue) solid dots, and the results corresponding to: the square root measurement, (black) crosses (which lie virtually on top of  the dots); the basic local strategy, (red) straight line; and the greedy strategy, (orange) triangles.}
\label{fig:1}
\end{figure}

Asymptotically, the maximum  success probability  is  attained by  the square root measurement \cite{pretty1,pretty2}.      Indeed, it is possible to show \cite{appendix} that the success probability  of the square root measurement, denoted by $P_{\rm SQ}$, satisfies the bound  
\begin{align}
P_{\rm SQ}  \ge   \left ( {\tr\sqrt G}/n  \right)^2
\end{align}
and therefore is equal  (at the leading order) to the  maximum success probability. 
  We also performed a  numerical analysis, revealing  that the square root  measurement is an 
 extremely good approximation already for short sequences ($n\gtrsim 10$), with a difference with respect to the optimal success probability
 of less than $10^{-3}$.          In Fig.~\ref{fig:1} 
we compare the asymptotic result in  Eq.~\eqref{uno} with the results for $n=50$ corresponding
 to the square root measurement and to the optimal measurement obtained via Semidefinite Programming~\cite{sdp}.  As it is apparent from the figure, the agreement is strikingly good.
The figure also includes the success probability of various local measurement strategies that will be discussed below.
 Notice that for $c=0$ the source states are orthogonal and perfect identification is possible, while in the limit  
$c\to1$  the source states  become indistinguishable and the success probability is given by random guessing, $P_{\max}=1/n\to 0$ for $n\to\infty$.  It is also patent from the figure that  Eq.~\eqref{uno} is a lower bound that becomes tight as $n$ goes to infinity, in agreement with the bounds (\ref{lower}) and (\ref{upper}). A numerical fit reveals that the correction to the leading order in Eq.~(\ref{uno}) is of order $1/n$, again, consistently  with our estimate.  

Both the optimal measurement and the square root measurement involve global operations on all the $n$ particles. This means that one has to scan the whole sequence of particles before getting an estimate of the change point.  We  now analyze the 
performance of online strategies, where each particle is individually measured as soon as it reaches the detector.    
The simplest such strategy  consists in measuring each  particle in the computational basis  $\{|0\rangle  , |1\rangle \}$.   The measurements are performed sequentially until the outcome $1$ is obtained for the first time, say at the $r$-th step. At this point, we will know for sure that the measured particle was in the state $|\phi\rangle$, meaning that the change must have  occurred  at some position $k\le r$. Our best guess for the change point is then $\hat k=r$, since this  
is the most likely hypothesis given the observed data. 
For the success probability one has the exact expression $P_{\mathrm{BL}}=1-c^2+c^2/n$, where BL stands for  ``basic local".  
%
This strategy is suboptimal for $0<c<1$, and remains suboptimal for $n\to\infty$.  
The relative difference between the two success probabilities can be of up to $50\%$ for suitable values of the overlap~$c$.
%

It is intriguing  to explore whether more general online strategies  can increase the success probability.  Let us consider a scenario where a classical learning  agent  is asked to guess   when the change point occurs.  The agent starts with a uniform prior $p(k)  =  1/n$ about  the location of the change point and updates her expectation as new data become available.      
  In order to update her information at the $s$-th step, the agent must perform a measurement, 
 which generally  depends on the  results, $r_1, r_2, \dots r_{s-1}$  obtained in the previous steps.  Here we focus on  \emph{greedy strategies}~\cite{greedy},~i.~e.~strategies that maximize the success probability at every step~\footnote{
In the considered scenario, the most general strategy can be optimized by dynamical programming, an optimization technique based on backward induction~\cite{bellman}.
Unfortunately, for the change point, the numerical overhead is prohibitive, as one needs to accurately approximate $n$-multidimensional functions. In practice, such technique is only suitable for problems with a limited number of hypothesis~\cite{julio-dynamical}.}. For these strategies, we determine the optimal measurement and the optimal guessing rule.     
The optimal strategy works as follows:      
At step $s$,  the agent has to perform  the Helstrom measurement \cite{helstrom} that distinguishes between the states $  |0\rangle$ and $|\phi \rangle$, given with (unnormalized) prior probabilities~\cite{appendix}
 \begin{align}
\nonumber  p^{(s)}_0  &:  =  \max_{k} \,  \{  p  (k|r_1,
 \dots, r_{s-1}) \}_{k=s+1}^n   \\
   p^{(s)}_\phi  &:  =  \max_{k} \,  \{  p  (k|r_1,
 \dots, r_{s-1}) \}_{k=1}^s \, .
 \end{align}
Here $p^{(s)}_0$ [$p^{(s)}_{\phi}$] is the probability of the most likely sequence that has the particle at position $s$ in the state $\ket{0}$ [$\ket{\phi}$]. . The agent can deduce these probabilities from the posterior probabilities, updated at step $s$. 
After the $s$-th measurement has been performed,  the prior is updated in accordance with the measurement result, using  Bayes' update rule:  adopting the shorthand notation $\eta^{(s)}_k   :  =   p(k|  r_1,\dots,  r_{s-1} )$, we have
\begin{equation}
\label{update-general}
\eta^{(s+1)}_{k} = \frac{p(r_s|k)  \,\eta^{(s)}_{k}}{\sum_{l=1}^n p(r_s|l)\,  \eta^{(s)}_{l}} \, . 
\end{equation}
%
After the last measurement, the agent updates  the prior to $\eta^{(n+1)}_{k}$ and produces the guess $\hat k$ that maximizes $\eta^{(n+1)}_{k}$ for the change point. 

For the  greedy strategy,  the full  optimization over the measurements has been  carried out analytically. However,  a direct quantification of the average performance is intractable, because the number of possible sequences of outcomes grows exponentially with $n$.  In order to compute the average success probability we used  a Monte Carlo simulation, leading to the values plotted in Fig.~\ref{fig:1} (orange triangles).  The figure shows  a significant enhancement of performance
over the basic local strategy.  Still,    the optimal greedy strategy does \emph{not} attain the optimal quantum performance, as the gap with the optimal collective strategy remains  even for large values of~$n$.  In short, this means that a learner with quantum memory outperforms a (greedy) learner with classical memory in the task of detecting change points. 

Having observed a gap between the optimal greedy strategy and the optimal quantum strategy, it is interesting to check whether the gap can be closed by using arbitrary local strategies, where each particle can be measured multiple times and the measurement settings can depend on the outcomes of all previous measurements.  Note that here the learner is allowed to use a quantum memory, but is limited to perform individual measurements on the particles. Unfortunately, optimizing over arbitrary local strategies is a daring task.    Nevertheless, we can provide an upper bound to the success probability by considering POVM operators that are positive under partial transposition (PPT). 
In this case, a numerical optimization proves the existence of a gap between   the local strategies  and the optimal collective strategy for every value of $n$ up to $n=7$.

In this Letter  we  introduced the quantum change point problem---a quantum version of the  problem of identifying  changes in a sequence of random variables.  In the quantum  change point problem,  a source emits particles in a default  state until a point where a mutation occurs, causing the source to switch to a different  state.  For pure states, we determined the maximum probability of correctly identifying the change point, showing that, for large sequences of particles, the optimal performance is attained by   the square root measurement. 
We also investigated  online strategies, where each particle is measured individually as soon as it is received from the source. Among the online strategies, we identified the optimal greedy strategy, which provides the best online guess  at each step.  Our calculations show  a gap between the  greedy strategy and the optimal quantum strategy based on a global measurement.  Further numerical optimization shows that the gap remains open even for arbitrary local strategies, indicating that local operations cannot match the performance of  the optimal quantum  protocol.   In particular, this shows that  a machine with  quantum memory  can outperform all machines with classical memory.

 \medskip 
 
\begin{acknowledgments}
\emph{Acknowledgments}. This research was supported by the Spanish MINECO
through contracts FIS2013-40627-P \& FIS2015-67161-P, the ERC Starting Grant 258647/GEDENTQOPT, the Generalitat
de Catalunya CIRIT contract 2014-SGR966,  the Foundational Questions Institute (FQXi- RFP3-1325 and FQXi-MGA-1502), the National Natural Science Foundation of China through Grant No. 11675136, the Hong Kong Research Grant Council through Grant No. 17326616, the Canadian Institute for Advanced Research, the Young 1000 Talents Program of China, and the HKU Seed Funding for Basic Research. 
\end{acknowledgments}

\bibliographystyle{unsrt}

\appendix

\section{SUPPLEMENTAL MATERIAL}

\section{Proof of theorem 1 and generalization to arbitrary priors}  

Here we provide a lower and an upper bound on the probability of correct state discrimination, valid for a generic set of linearly independent pure states $\{   |\Psi_k\rangle\}_{k=1}^n$  and for a generic choice of prior probabilities  $\{  p_k\}_{k=1}^n$.   
The bounds are expressed in terms of the Gram matrix of the weighted states 
\begin{align}
|\widetilde \Psi_k\rangle :  =  \sqrt  {p_k}\, |\Psi_k\rangle  \, ,
\end{align}
that is, the matrix $W$ with elements 
\begin{align}W_{ij}  :  =  \langle \widetilde \Psi_i|  \widetilde \Psi_j\rangle   \, .
\end{align}

The maximum success probability can be estimated with the following Theorem, which generalizes Theorem~1 of the main text to arbitrary prior distributions:  
\begin{theo}
Let $\{  |\Psi_k\rangle\}_{k=1}^n$ be a set of linearly independent pure states. 
The maximum probability of correctly identifying a state drawn from the set  $\{  |\Psi_k\rangle\}_{k=1}^n$ with probability $\{p_k\}_{k=1}^n$ satisfies the bounds 
\begin{align}\label{app:lower}
 P_{\max} \ge    \frac{\left(   {\tr \sqrt W}\right)^2 } n
 \end{align}
and 
\begin{align}\label{app:upper} 
P_{\max} \le     \frac{\left(   {\tr \sqrt W}\right)^2 } n      +     \sqrt{n\, \lambda_{\max }}    \,  \| {\bf q}  -  {\bf u} \|_1  \, , 
\end{align}  
where   $\lambda_{\max}$ is the maximum eigenvalue of $W$,     \mbox{${\bf q}   =  \{q_k\}$} is the probability distribution defined by  \mbox{$q_k  :  =   \big(  \sqrt W\,\big)_{kk}/\tr\big(\sqrt W\,\big)$}, $  {\bf  u}  =  \{u_k\}$ is the uniform distribution,  and $\|  {\bf q}  - {\bf u}\|_1  : =\sum_k  |q_k-u_k|$ is the trace norm. 
 \end{theo}

{\bf Proof.}    Since the states  $\{  |\Psi_k\rangle\}_{k=1}^n$ are linearly independent, the optimal measurement consists of orthogonal rank-one projectors~\cite{belavkin}.  Let us denote the projectors by $M_k  =  |m_k\rangle\langle m_k|$, for a suitable orthonormal basis $\{  |m_k\rangle\}_{k=1}^n$.  Then,  the probability of correct discrimination can be written as  
\begin{align}
\nonumber P_{\rm succ}   &=   \sum_k  p_k \,   |\langle m_k|  \Psi_k\rangle|^2  \\  
\nonumber &  =  \sum_k      |\langle m_k|  \widetilde \Psi_k\rangle|^2\\ 
&   =   \sum_k    \left|   B_{kk} \right|^2\, ,
\end{align}  
 where $B$ is the matrix defined  by  the relation
  \begin{align}  |\widetilde \Psi_k \rangle  =   \sum_i  \,   B_{ik} \, |m_i \rangle \, .
 \end{align}
  By definition, one has  $B^\dag B  =  W$. 
      Hence, the  polar decomposition yields the relation  
      \begin{align}\label{polar} 
      B   =   U \sqrt W \, ,
      \end{align}
      for a suitable unitary matrix $U$.    
          Note that a generic change of orthonormal basis,  
  \begin{align}
 |m_i\rangle \to   |m_i'  \rangle     =    \sum_j   V_{ji}  \,  |m_j\rangle 
  \end{align}   
(where $V$ is a generic unitary matrix), results into the change of matrix
  \begin{align}\label{changeV}
    B \to   B'  =   V^\dag   B \, .
  \end{align}     
  Combining Eqs. (\ref{polar}) and (\ref{changeV}),   the maximum probability of correct discrimination can be expressed as  
\begin{align}\label{pmax}
 P_{\max} =    \max_U  \,  \sum_k   \left|  \left(   U   \sqrt W \right  )_{kk} \right|^2 \, .
\end{align}    
 Setting $U=\openone$, one has the lower bound  
\begin{align}
P_{\max}  \ge    \sum_k     \left(  \sqrt W \right  )^2_{kk}   \ge  \frac{\left(  \tr \sqrt W  \right)^2}n  \, ,
\end{align}
the second inequality following from the  convexity of the function $f(x)= x^2$.    This proves the lower bound  (\ref{app:lower}).

  Let us prove the upper bound (\ref{app:upper}).  Using the Cauchy-Schwarz inequality in Eq.~(\ref{pmax}), we obtain the upper bound 
 \begin{align*}
 P_{\max}   & =   \max_U  \,  \sum_{k}   \left|  \sum_{s}\left(U  W^{\frac{1}{4}}\right)_{ks} \left(W^{\frac{1}{4}}\right)_{sk} \right|^2 \\
& \le     \max_U  \,  \sum_k     \left( \sqrt W\right)_{kk} \,   \left(   U   \sqrt W   U^\dag  \right)_{kk}  \\
    &  =     \tr \left(\sqrt W\right)    \times  \,   \max_U  \left[ \sum_k     q_k       \,   \left(  U   \sqrt W   U^\dag  \right)_{kk}   \right]   \, . 
\end{align*}    
Moreover, the argument of the maximum can be upper bounded as  
\begin{align*}
 \sum_k     q_k       \,   \left(  U   \sqrt W   U^\dag  \right)_{kk}       &\le  \sum_k      \frac 1n \,         \left(  U   \sqrt W   U^\dag  \right)_{kk}      \\
   & \quad +      \sum_k      \left|q_k  -   \frac 1n \right|       \left(  U   \sqrt W   U^\dag  \right)_{kk}     \\
   &   \le   \frac {  \tr  \sqrt W }n   +   \|  {\bf q}-  {\bf u} \|_1  \,   \sqrt{ \lambda_{\max} }\, .    
\end{align*}
Finally, from Eq. (\ref{app:lower}) we have the bound  
\begin{align}
\tr \sqrt W   \le \sqrt {n  \,   P_{\max}}  \le \sqrt n  \, .  
\end{align}
Combining the above inequalities we obtain the desired upper bound (\ref{app:upper}). $\blacksquare$

\medskip  

When the prior distribution is uniform, the weighted Gram matrix $W$ is given by $W=  G/n$, where $G$ is the unweighted Gram matrix used in the main text. Substituting this relation into the bound (\ref{app:upper}) one obtains Eq.~(5) of the main text.

\section {\boldmath Eigenvalues and eigenvectors of $H$}

In this section we derive explicit expressions for the eigenvectors and eigenvalues of the matrix $H$.   

To this purpose, it is useful to  first recall some properties of 
the Chebyshev polynomials of the second kind, denoted by $U_n(x)$.  The Chebyshev polynomials of the second kind  can be defined as the characteristic polynomial of the tridiagonal matrix  $T$ of size~$n$ whose entries are $T_{ij}=\delta_{i\,j+1}+\delta_{j\,i+1}$. Specifically, \mbox{$U_n(x)=\det(2x\,\openone-T)$}, i.e., the eigenvalues of~$T$ are defined to be twice the roots, $x_l$, of~$U_n(x)$. By expanding the determinant by the first row one readily obtains the well known recursion relation $U_n(x)=2x \,U_{n-1}(x)-U_{n-2}(x)$~\cite{abramowitz}. One can check that this recursion relation along with the initial conditions in standard form, $U_0(x)=1$, $U_{-1}(x)=0$, give the right characteristic polynomial for {\em any} size of $T$. It suffices to check the~$n=1,2$ cases. One has~$U_1(x)=2x$ and  $U_2(x)=4x^2-1$, which are indeed the characteristic polynomials of $T$ of sizes 1 and~2.  

We now turn to the eigenvalues and eigenvectors of~$H$, which we will compute using a different approach. The matrix $H$ is nothing but the matrix $T$ with the addition of two extra entries at each end of the principal diagonal, namely,
$H_{ij}= T_{ij}+c\,( \delta_{i \,1}\delta_{j\,1}+ \,\delta_{i\, n}\delta_{j\, n})$. Let us denote by~$2x_l$ the $l$-th eigenvalue of $H$ and by $\vec{w}^{\, l}$ the corresponding
unnormalized eigenvector, chosen with the convention~$w^l_1=1$. The equation $H\vec{w}^{\, l}=2x_l \vec{w}^{\, l}$ 
is equivalent to the following system of linear equations: 
\begin{eqnarray}
\label{eigenvectors}
c w_1^l+w^l_2&=& 2x_l\, w_1^l; 
\nonumber \\
w^l_{j-1}+w^l_{j+1} &=& 2x_l\, w_j ^l, \qquad  2\le j\le n-1; \\
 w_{n-1}^l+ c w^l_{n}&=& 2x_l\, w_n^l. 
\nonumber
\end{eqnarray}
The second line of this system can be viewed as the recursion relation $w^l_{j+1} = 2x_l\, w_j ^l - w^l_{j-1}$, 
which is the recursion relation of the Chebyshev polynomials given above (with $n\to j+1$). It follows that the first and second line of Eq.~(\ref{eigenvectors}), along with the convention $w^l_1=1$, imply
\begin{equation}
w^l_j=U_{j-1}(x_l)-c\,U_{j-2}(x_l),\qquad j=1,2,\dots, n.
\label{wl}
\end{equation}
Since all the components of $\vec{w}^{\, l}$ have been determined, the third line in Eq.~(\ref{eigenvectors}) must give the eigenvalues of~$H$.
By substituting Eq.~(\ref{wl}) in the third line of Eq.~(\ref{eigenvectors}) and using the Chebyshev recursion relation again, one obtains 
\begin{equation}
0=U_n(x_l)-2c\,U_{n-1}(x_l)+c^2 U_{n-2}(x_l):=P_n(x_l),
\label{P_n again}
\end{equation}
which must hold for $l=1,\dots,n$.  The polynomial~$P_n(x)$ has degree $n$ and its $n$ roots, $x_l$, give the eigenvalues of $H$ as $2x_l$. Note that $P_n(x)$ has to be proportional to the characteristic polynomial of $H$, i.e., $P_n(x)\propto \det(2x\openone-H)$, as both polynomials have the same degree and the same zeroes. 

\section {\boldmath Distribution of the eigenvalues of $H$}

Here   we  analyze the distribution of the zeroes of the polynomial  $P_n(x)$ defined in Eq. (\ref{P_n again}).  

Setting $x=  \cos \theta$, we recall that the Chebyshev polynomial $U_n (\cos \theta)$ can be expressed as \cite{abramowitz}
\begin{equation}
U_n(\cos\theta)={\sin(n+1)\theta\over \sin\theta} \, .
\label{U_n(cos)}
\end{equation}
Then, a little bit of trigonometry yields the relation 
\begin{equation}
P_n(\cos\theta)=A(\theta)\sin\left[n\theta+\delta(\theta)\right],
\label{beat}
\end{equation}
with
\begin{eqnarray}
\left  \{  
\begin{array}{lll}    A(\theta)&:=&\displaystyle{1-2c\cos\theta+c^2\over\sin\theta}\\
&&\\
\delta(\theta)&:=&\displaystyle\arctan{(1-c^2)\sin\theta\over(1+c^2)\cos\theta-2c}
\end{array}
\right.   \, ,
\end{eqnarray}
%
where $0<\delta(\theta)<\pi$.
{}From Eq. (\ref{beat}) we can see  that every zero of $P_n(\cos\theta)$ must be the solution to one of the equations 
\begin{align} n\theta+\delta(\theta)=  l \pi   \, , \qquad l=1,2,\dots,n \, .
\end{align}  
Denote by $\theta_l$ the angle that solves the $l$-th equation.   Since $\delta (\theta_l)$ is contained in the interval $(0,\pi)$, we have the bound
\begin{equation}
{\pi \over n} l \le \theta_l \le {\pi \over n}  \left(l+1\right)  \, .
\end{equation}
In other words, the interval $(0,\pi)$ can be divided into intervals of length $\pi/n$, with the $l$-th interval containing the zero  $\theta_l$.  For large $n$, this means that the zeros are uniformly distributed in the interval $(0,\pi)$.

\section{\boldmath  The trace and diagonal matrix elements of $\sqrt G$}

Here we evaluate the normalized trace  $\sqrt G/n$ and we quantify its deviation from the limit value $\gamma:  =  \lim_{n\to\infty  }  \tr \sqrt G /  n$.   In the process of computing the trace, we also evaluate the diagonal matrix elements of $\sqrt G$, which will become useful in the next section.    

We proceed along the following steps:  
\begin{enumerate}
\item construct the normalized eigenvectors of $\sqrt G$
\item evaluate the diagonal elements
\item evaluate the trace.
\end{enumerate} 

\subsection{The normalized eigenvectors of $\sqrt G$}  

The eigenvectors of $\sqrt G$ coincide with the eigenvectors of the matrix $H$, provided in Eq. (\ref{wl}). 
Setting  $x_l  =  \cos \theta_l$, we can use the trigonometric representation of  the Chebyshev polynomials given in   Eq.~(\ref{U_n(cos)}).   In this way, we obtain  
\begin{equation}
{w}^l_j={\sin( j\theta_l)-c \sin[(j-1)\theta_l]\over \sin\theta_l}.
\end{equation}
Now, the norm  $\vec w^l$ can be evaluated explicitly as 

\begin{align}
\|\vec w^l\|^2&:= \sum_j  |  w^l_j|^2 \nonumber   \\
& ={n\over2\sin^2\theta_l}
\left\{
1-2c \cos\theta_l+c^2\phantom{{\sin(2n\theta_l)\over2n\sin\theta_l}}\right.\nonumber\\
& \quad +{1-c^2\over2n}\left[1-\cos(2n\theta_l)\right]\nonumber\\
&\quad- \left. {\sin(2n\theta_l)\over2n\sin\theta_l}\left[(1+c^2)\cos\theta_l-2c\right] \right\}  \\
  &    = {n\over2\sin^2\theta_l}
\left\{
1-2c \cos\theta_l+c^2      +   \frac{ f_n(\theta_l)}n  \right\}       \, , \nonumber 
\end{align}
having defined the function
\begin{align}
\nonumber  f_n(x)    &:  = {1-c^2\over2}\left[1-\cos(2n x)\right]   \\ 
&  \quad  -   {\sin(2n x)\over2\sin x}\left[(1+c^2)\cos x -2c\right] \, .
 \end{align}
 Defining      the normalized eigenvectors $\vec v^l:=\vec w^l/\|\vec w^l\|$, we then have 
\begin{align}\label{vl}
\left| v^l_j  \right|^2    = \frac 2 n \frac{   \left [\,  \sin( j\theta_l)-c \sin(j-1)\theta_l\, \right]^2      }{   1-2c \cos\theta_l+c^2      +    f_n(\theta_l)/n}  \, .
\end{align}

\subsection{The diagonal elements of $\sqrt G$}

Having computed the eigenvalues and eigenvectors of the Gram matrix $G$, we can now evaluate the diagonal elements of its square root $\sqrt G$.  
  We start from the expression 
  \begin{align}
  \sqrt G  = \sum_l   \sqrt {\lambda_l} \,   |v^l\rangle\langle v^l| \, ,
  \end{align} recalling that the eigenvalues are given by   
  \begin{align}\label{lambdal}\lambda_l  =  {\frac{1-c^2}{ 1-2c \cos\theta_l+c^2}} \,  .
  \end{align} 
   Then, the diagonal elements of $\sqrt G$ are 
  \begin{align}
  \left(  \sqrt G\right)_{kk}  &   =  \sum_l     \sqrt{\lambda_l}  \,  |v^l_k|^2  \, ,  
  \end{align}
   with $v^l_k$ given as in Eq. (\ref{vl}).   Explicitly, the matrix element $\big(\sqrt G\,\big)_{kk}$   is given by 
\begin{align}
  \nonumber \left(  \sqrt G\right)_{kk}      
  \nonumber &   =    \frac 1 n \,   \sum_l \sqrt{  {\frac{1-c^2}{ 1-2c \cos\theta_l+c^2}}}     \\  
    \label{sqrtGkk}&  \qquad \times \,          {{\left[ \sin k\theta_l-c\sin(k-1)\theta_l\right]^2}\over { 1-2c \cos\theta_l+c^2     +     f_n(\theta_l  )/n }}   
\end{align}

   We now show that most of the matrix elements $\big(\sqrt G\,\big)_{kk}$ are approximately equal to  the   limit value 
  $  \gamma:    =\lim_{n\to \infty}  \tr \sqrt G /n$.  
    Note that $\gamma$ can be computed explicitly in terms  the eigenvalues: indeed, one has  
   \begin{align} \nonumber 
   \gamma & =   \lim_{n\to \infty}  \frac 1 n \,  \sum_l    \sqrt \lambda_l \\
  \nonumber  & =   \lim_{n\to \infty}  \frac 1 n \,  \sum_l      \sqrt{\frac{1-c^2}{ 1-2c \cos\theta_l+c^2}}  \\
   &    =     \frac 1 \pi  \, \int_0^\pi {\rm d} \theta\,  \sqrt{\frac{1-c^2}{ 1-2c \cos\theta+c^2}} \, .
   \end{align}  
   
\begin{figure}[bhtp] 
$$
   \includegraphics[width=27em]{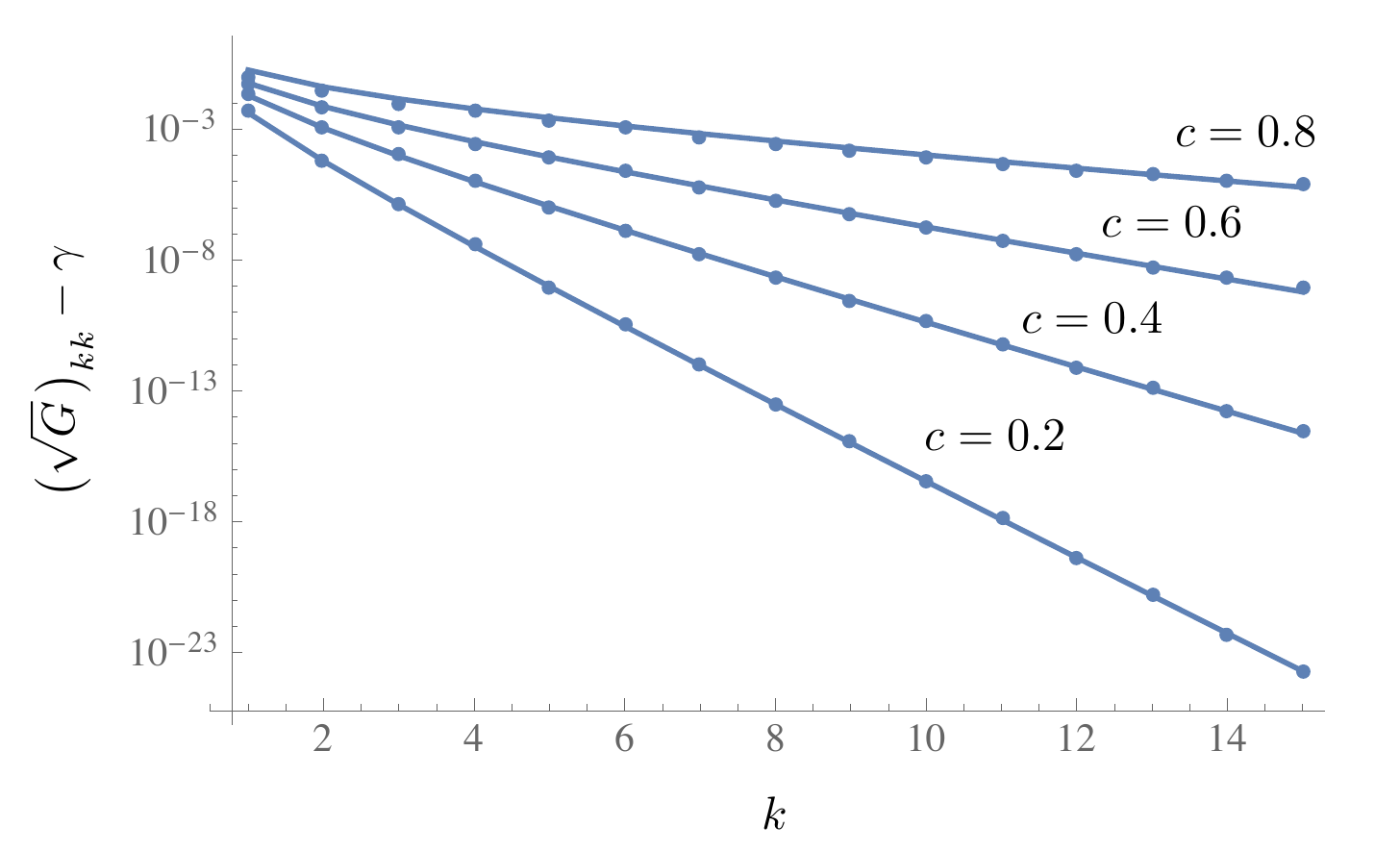} 
 $$
\caption{  Log-scale plots of the deviation $\big(    \sqrt G\,\big)_{kk}-\gamma$, for  $n=30$ and $k$ varying from 1 to 15.  The solid lines are the asymptotic approximation on r.h.s of Eq.~(\ref{skkminusave}) and the dots are the result of numerical evaluation of $\sqrt G$ from $G$.}
\label{fig:3}
\end{figure}

We now show that the deviation vanishes for all values of $k$ in the interval $[ n^\epsilon   ,   n- n^\epsilon]$.    
 To this purpose, we evaluate the matrix element  $\big( \sqrt G\,\big)_{kk}$ at the leading order of the large $n$ asymptotics,  obtained  by replacing the sum  in Eq. (\ref{sqrtGkk}) by an integral and by dropping the term~$f_n(  \theta_l)/n$ in the denominator. In this way, we  obtain the approximate equality
\begin{align}\label{approx}
   \left(  \sqrt G\right)_{\!kk} \!\!   \approx  \frac {\sqrt{1\!-\!c^2}}\pi   \! \int_0^\pi \!\! {\rm d} \theta      \,         {{\left[ \sin k\theta-c\sin(k-1)\theta \right]^2}\over {   (1-2c \cos\theta +c^2 )^{3/2}    }}    \, .
\end{align}
 Then,  some elementary algebra  gives 
%
\begin{equation}
\left(\sqrt G\right)_{kk}-\gamma \approx
{\sqrt{1\!-\!c^2}\over\pi}\left( 2c I_{2k-1}\!-\!I_{2k}\!-\!c^2 I_{2k-2}\right) \, ,
\label{deviation}
\end{equation}
 where the integrals $I_r$ are defined as
\begin{equation}
I_r\!  :=\!\int_0^\pi\! {\cos r\theta\; d\theta\over\left(1-2c \cos\theta+c^2\right)^{3/2}} \, .
\label{I_m}
\end{equation}
We then show that  the integrals $I_r$ vanish exponentially with $r$:  
\begin{prop}\label{prop:integral}
For $c<1$,  the leading order of the integral $I_r$ in Eq. (\ref{I_m}) is given by
\begin{equation}
I_r={2 c^r\sqrt{\pi r}\over(1-c^2)^{3/2}} \, .
\label{I_m asympt}
\end{equation}
\end{prop}
The proof can be found in the end of this subsection.   Inserting the asymptotic expression (\ref{I_m asympt}) into Eq. (\ref{deviation}) we  obtain the relation
\begin{equation}\label{skkminusave}
\left(  \sqrt G\right)_{kk}\!-\gamma \approx    {1   \over  {4 (1-c^2)}} ~ {{c^{2k}}   \over  {\sqrt{2\pi  k^3}}} \, , 
\end{equation}
valid in the interval $[  n^\epsilon  ,  n-  n^\epsilon]$.   In conclusion, the deviation   $\big(    \sqrt G\,\big)_{kk} - \gamma$  decays exponentially with $k$.   
We stress that the error introduced by the approximation  (\ref{approx}) is negligible with respect to  the leading order,  quantified by the r.h.s. of Eq. (\ref{skkminusave}).    This point is illustrated in Fig.~\ref{fig:3}, which compares the  the r.h.s. of Eq. (\ref{skkminusave}) with  the exact values of the deviation, computed by direct numerical evaluation of $\sqrt G$ from $G$. Log-scale plots are shown for various values of the overlap~$c$, setting $n=30$ and letting~$k$ vary from 1 to 15. 
The agreement is extremely good and  backs up the validity of the approximation (\ref{approx}) even for small values of $k$.   

 In the next subsection we will use Eq. (\ref{skkminusave})  to quantify the deviation between  the diagonal  of the matrix $\sqrt G/\tr(\sqrt G)$ and  the uniform distribution.     

\medskip 

{\bf Proof of Proposition \ref{prop:integral}.} 
We start by noticing that the integral on the r.h.s of Eq.~(\ref{I_m}) can be expressed as a contour integral over the unit circle~$C$ on the complex plane:
\begin{equation}
I_r=
{1\over2i}\oint_C dz\;{z^{r+1/2}\over\left[z-c(z^2+1)+c^2 z\right]^{3/2}} .
\label{I_m complex}
\end{equation}

\begin{figure}[bhtp] 
$$
   \includegraphics[width=2.1in]{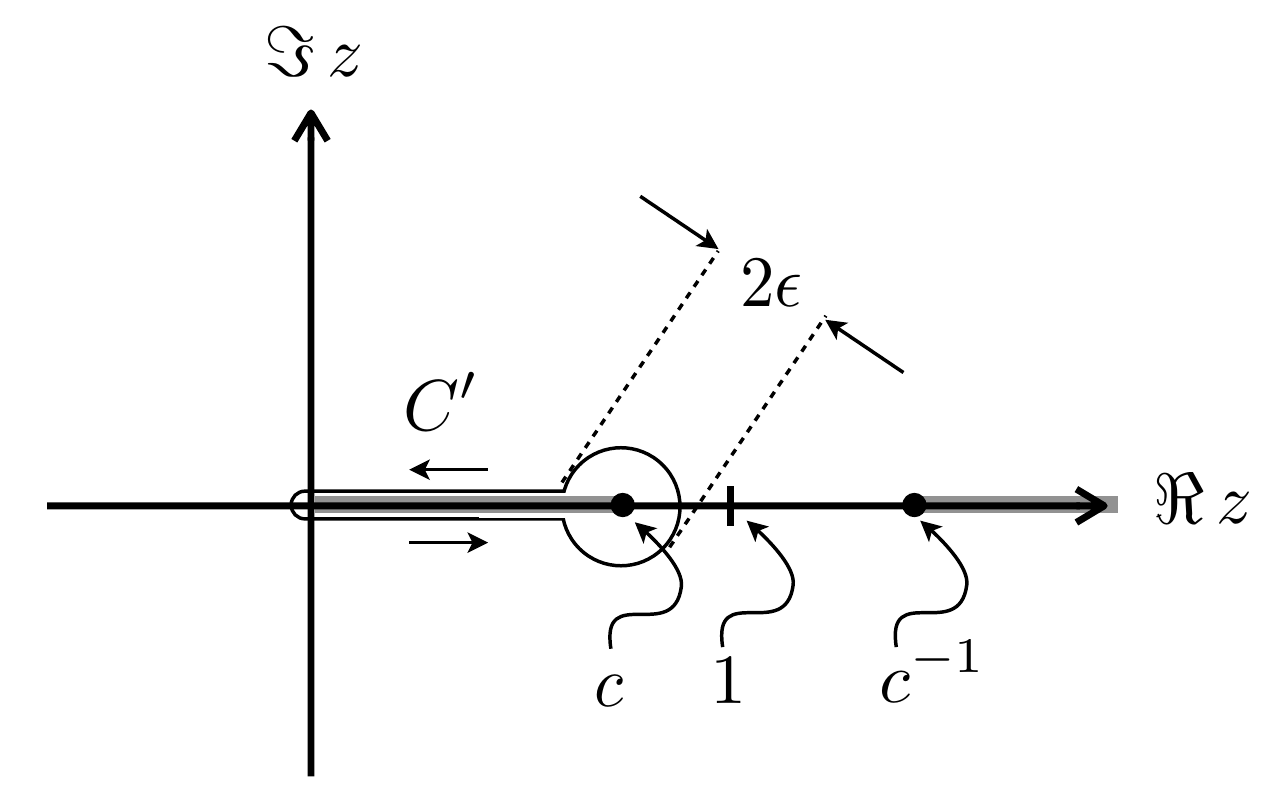} 
 $$
\caption{The figure shows (gray) the branch cuts of the integrand in Eq.~(\ref{I_m complex}) and the contour $C'$ used to obtain Eq.~(\ref{I_m 2 int}).}
\label{fig:2}
\end{figure}

We can choose the branch of the integrand so that its branch cuts are the intervals~$[0,c]$ and~$[c^{-1},\infty)$ on the real axis. Since this branch is analytic elsewhere~and the cut~$[c^{-1},\infty)$ is outside $C$, we can deform the contour~$C$ to a new contour $C'$ around $[0,c]$. One readily sees that the integrand in Eq.~(\ref{I_m complex}) behaves as $(z-c)^{-3/2}$, so care must be taken to evaluate the new contour integral near the end point $z=c$, as some divergencies may arise because of this singular behavior. Specifically, we choose~$C'$ as in Fig.~\ref{fig:2}, where $\epsilon>0$ and the limit $\epsilon\to0$ is implicit. As a result, $I_r$ has a contribution coming from the discontinuity of the integrand along the interval $[0,c-\epsilon]$ and a contribution coming from the integration around the circle $C_\epsilon$ of radius $\epsilon$ and center at $z=c$:
\begin{eqnarray}
I_r
&=&-{1\over c^{3/2}}\int_{0}^{c-\epsilon}dx\;{x^{r+1/2}\over\left\{(c-x)[(1/c)-x]\right\}^{3/2}}\nonumber\\
&+&{1\over2ic^{3/2}}\oint_{C_\epsilon}dz{c^{r+1/2}\over(z-c)^{3/2}[(1/c)-c]^{3/2}}.
\label{I_m 2 int}
\end{eqnarray}
Note that the limit $\epsilon\to0$ of each separate line is ill-defined, as they both diverge as $\epsilon^{-1/2}$. To circumvent this problem,  
we write $(c-x)^{-3/2}=2(d/dx)(c-x)^{-1/2}$ and integrate the first line of~Eq.~(\ref{I_m 2 int}) by parts. In doing so, we see that the $\epsilon^{-1/2}$ terms cancel and we obtain the simple expression 
\begin{equation}
I_r=2\int_{0}^{c}{dx\over (c-x)^{1/2}}{d\over dx}{x^{r+1/2}\over[1- c x]^{3/2}}.
\end{equation}
We can further simplify this expression using the change of variable $x= c t$, which enables us to express $I_r$ in terms of hypergeometric functions. However, we are just interested in the asymptotic behavior of $I_r$. Keeping only the leading contribution as $r$ goes to infinity, we have
\begin{equation}
I_r=2r c^r\int_0^1 dt\, t^{r-1/2}(1-t)^{-1/2}(1-c^2 t)^{-3/2}. 
\label{I_m t-int}
\end{equation}
The asymptotic behaviour of this integral can be easily evaluated by noticing that the leading contribution comes from the region near the upper limit of integration, so we can set~$t=1$ in the last factor in Eq.~(\ref{I_m t-int}) and write
\begin{eqnarray}
I_r\!\!&=&\!\!{2rc^r\over(1-c^2)^{3/2}}\int_0^1\!\! dt\, t^{r-1/2}(1-t)^{-1/2}\nonumber\\[.5em]
\!\!&=&\!\!{2rc^r B(\mbox{${1\over2}$},r\!+\!\mbox{${1\over2}$})\over(1\!-\!c^2)^{3/2}}\nonumber\\[.5em]
&=&{2\sqrt\pi c^r\,\Gamma(r\!+\!\mbox{${1\over2}$})\over (r\!-\!1)!(1\!-\!c^2)^{3/2}},
\label{I_m Gamma}
\end{eqnarray}
where $B(a,b)$ is the Euler Beta function,
\begin{eqnarray}
B(a,b)=\int_0^1 dt\; t^{a-1}(1-t)^{b-1},\\[-.1em]\nonumber
\end{eqnarray} 
and we have used the relation,
\begin{equation}
B(a,b)={\Gamma(a)\Gamma(b)\over\Gamma(a+b)}.
\end{equation}
Using the Stirling formula in the third line of Eq.~(\ref{I_m Gamma}), we finally obtain Eq.~(\ref{I_m asympt}). $\blacksquare$

\subsection{The trace of $\sqrt G$}  
 Here we show that the normalized trace $\tr \sqrt G/n$ is close to its limit value $\gamma$, up to an error of size $1/n^{1-\epsilon}$, where $\epsilon$ is an arbitrary constant in the interval $(0,1)$.  
    For this purpose, we divide the values of $k$ into two subsets, defined as 
\begin{align}
\nonumber {\mathsf S}   &:  =  \{  k   \in \mathbb N :       \lceil n^{\epsilon}  \rceil \le k   \le n-\lceil n^{\epsilon}  \rceil \     \}    \\
\label{sets} \overline{\mathsf S} & :  =  \{  1,\dots,  n\}\setminus \mathsf S \, .  
\end{align}  
 
 The trace of $\sqrt G$ can be evaluated as  
\begin{align}
\nonumber  \tr \sqrt G   &  =    \sum_{k  \in  \mathsf S}   \, \left(  \sqrt G\right)_{kk}   +   \sum_{k  \in  \overline {\mathsf S}}   \, \left(  \sqrt G\right)_{kk}  \\
\nonumber  &  =         \sum_{k  \in  \mathsf S}   \, \left(  \gamma  +    {1   \over  {4 (1-c^2)}} ~ {{c^{2k}}   \over  {\sqrt{2\pi  k^3}}}  \right)      +    \sum_{k  \in  \overline {\mathsf S}}   \, \left(  \sqrt G\right)_{kk} \\
&  =        | \mathsf{S}  \,   |  \gamma  +    \sum_{k  \in  \overline {\mathsf S}}   \, \left(  \sqrt G\right)_{kk}   +    O \left(  n^{1-3\epsilon/2}c^{2n^\epsilon}\right)   \, ,  
\end{align}
the second equality following from  Eq. (\ref{skkminusave}).  

Using the above expression it is easy to produce upper and lower bounds on  $\tr \sqrt G$.  An upper bound is obtained as follows:
\begin{align}\label{upperboundtrsqrt}
\nonumber\tr\sqrt G  & \le   | \mathsf{S}|    \, \gamma       +  \sqrt{  \lambda_{\max}} \,  |\overline {\mathsf S}|   +     O \left(  n^{1-3\epsilon/2}c^{2n^\epsilon}\right)  \\
   &   \le     n \,  \gamma    +     \sqrt{\lambda_{\max}}       |\overline {\mathsf S}|   +     O \left(  n^{1-3\epsilon/2}c^{2n^\epsilon}\right)       \, .    
\end{align} 
Similarly, we have the lower bound  
\begin{align}
\nonumber\tr\sqrt G  & \ge   | \mathsf{S}|    \, \gamma  +     O \left(  n^{1-3\epsilon/2}c^{2n^\epsilon}\right)  \\
 &  =      n \,  \gamma    -  \gamma  |\overline {\mathsf S}|   +     O \left(  n^{1-3\epsilon/2}c^{2n^\epsilon}\right)     \, .    
\end{align} 
Using the relation   $  2  n^{\epsilon}  \le |\overline {\mathsf S}|  \le 2 \, (n^{\epsilon}+1)  $ we finally obtain 
 the bounds  
\begin{align}
\gamma    -   \frac{   2\gamma}{n^{1-\epsilon}}   \le  \frac{\tr \sqrt G}n \le   \gamma +   2  \sqrt{\lambda_{\max}}  \left( \frac{   1}{n^{1-\epsilon}}  +  \frac 1n  \right)    \, ,
\end{align}
valid up to an exponentially small correction of size~$   O \left(  n^{-3\epsilon/2}c^{2n^\epsilon}\right)$.  
 More compactly, the above bounds can be written as 
 \begin{align} 
   \left|  \frac{  \tr \sqrt G} n  -  \gamma  \right|  \le          \frac { 2 \, \max  \{ \gamma, \sqrt{ \lambda_{\max}}\}  }{n^{1-\epsilon}}    +       O \left(   \frac 1n\right)  \, .
 \end{align}
Now, direct inspection shows that $\sqrt{\lambda_{\max}}$ is always larger than $\gamma$. Hence,  the bound becomes  
 \begin{align}\label{normtracebound}
    \left|  \frac{  \tr \sqrt G} n  -  \gamma  \right|  \le          \frac { 2 \sqrt{ \lambda_{\max}} }{n^{1-\epsilon}}    +       O \left(    \frac 1n\right)  \, .
 \end{align}

\section{\boldmath Deviation of ${\bf q}$ from the uniform distribution}
In this section we consider  the probability distribution $  {\bf q}   =  \{   \big( \sqrt G \, \big)_{kk}/\tr \sqrt G\}$  for the change point problem and we quantify the deviation of $ \bf q$ from the uniform distribution.

Here we upper bound the trace distance between the probability distribution ${\bf q }=\{   q_k\} $ defined by  
\[  q_k  :  =  \frac{ \left( \sqrt G \right)_{kk}}{\tr \sqrt G}\] and the uniform distribution, denoted by $\bf u$.    Our strategy is to separately analyze the contributions to the trace distance coming from the two sets $\mathsf S$ and $\overline  {\mathsf  S}$   defined in Eq.~(\ref{sets}).

Let us consider first the contribution of the set $\mathsf S$.  For~$k\in\mathsf S$, we have  
\begin{align}
\nonumber 
\left|  q_k  -  \frac 1n  \right|   &     =  \left|     \frac {  \left(  \sqrt G\right)_{kk}    -   \tr \sqrt G  /n} {  \tr \sqrt G} \right|  \\
  &     \le      \frac {  \left|  \left(  \sqrt G\right)_{kk}    - \gamma  \right|  +  \left|   \gamma  -  \tr \sqrt G  /n  \right| } {  \tr \sqrt G}   \, .
\end{align}
Now, the first term is upper bounded as 
\begin{align}
 \nonumber \left|  \left(  \sqrt G\right)_{kk}    - \gamma  \right|    &  \le    {1   \over  {4 (1-c^2)}} ~ {{c^{2n^\epsilon}}   \over  {\sqrt{2\pi  n^{3\epsilon}}}} \\
\label{primo}  &  =   O \left(  n^{-3\epsilon/2}c^{2n^\epsilon}\right)     \, ,
\end{align}  due to  Eq. (\ref{skkminusave}). The second term is upper bounded by Eq. (\ref{normtracebound}). 
Hence, the contribution of $\mathsf S$ to the trace distance can be upper bounded as  
 \begin{align}
\nonumber 
\sum_{k\in\mathsf S}    \left|  q_k  -  \frac 1n  \right|   &  \le  \sum_{k\in\mathsf S}            \frac{       { 2 \sqrt{ \lambda_{\max}} }/{n^{1-\epsilon}}    +       O \left( 1/n \right)    } {\tr \sqrt G}  \\
 \nonumber 
  & \le  \sum_{k\in\mathsf S}            \frac{        { 2 \sqrt{ \lambda_{\max}} }/{n^{1-\epsilon}}  +         O \left( 1/n\right) } {n\sqrt {\lambda_{\min}}}  \\
\nonumber 
&  \le              \sqrt{\frac{  \lambda_{\max}} {\lambda_{\min}}}        \, \frac{2}{n^{1-\epsilon}}  +           O \left( \frac 1n\right)
 \, , \label{vanish1}
 \end{align}
 having used the relation  
 \begin{align}
 \tr \sqrt G   \ge n    \,    \sqrt{\lambda_{\min}}  \, , 
 \end{align}
 where $\lambda_{\min}$ is the minimum eigenvalue of $G$.  
  In conclusion, Eq. (\ref{vanish1}) shows that the contribution of the set $\mathsf S$ vanishes     in the large $n$ limit.  
 
Let us consider the contribution of the set $\overline {\mathsf S}$.  For $k\in  \overline{\mathsf S}$, we have  the inequality
\begin{align}
\nonumber 
\left|  q_k  -  \frac 1n  \right|   &     =  \left|     \frac {  \left(  \sqrt G\right)_{kk}    -   \tr \sqrt G  /n} {  \tr \sqrt G} \right|  \\
\nonumber &  \le    {  \sqrt{\lambda_{\max}}    \over  { \tr \sqrt G  }}   \\
  & \le    \sqrt{\frac{\lambda_{\max}}{\lambda_{\min}}}   \,  \frac 1n  \, , 
\end{align}
 which  leads to  the upper bound  
\begin{align}
\label{vanish2}   \sum_{k\in   \overline{\mathsf S}}   \left|  q_k  -  \frac 1n  \right|   &   
 \le    \sqrt{\frac{\lambda_{\max}}{\lambda_{\min}}}   \, \frac2  {n^{1-\epsilon}}     \, . 
\end{align}

Using the bounds (\ref{vanish1}) and (\ref{vanish2}), the deviation between~$\bf q$ and the uniform distribution can be upper bounded as  
\begin{align}
\nonumber   \|  {\bf q }  -  {\bf u}\|_1    &  =  \sum_{k\in    {\mathsf S}}   \left|  q_k  -  \frac 1n  \right|   +    \sum_{k\in   \overline{\mathsf S}}   \left|  q_k  -  \frac 1n  \right|    \\
\nonumber &  \le 
      \sqrt{\frac{  \lambda_{\max}} {\lambda_{\min}}}        \, \frac{4}{n^{1-\epsilon}}  +           O \left( \frac1n\right)  \\
      &   \le 
       {\frac{1+c} {1-c}}        \, \frac{4}{n^{1-\epsilon}}  +           O \left( \frac 1n\right) \, , 
\end{align}
having used the bounds  
\begin{align}  \lambda_{\max}  \le (1+c)/(1-c)
\end{align} and 
\begin{align}\lambda_{\min}  \ge (1-c)/(1+c) \, ,
\end{align} following from Eq. (\ref{lambdal}).  

\section{Lower bound on the success probability of the square root measurement}

For  a generic set of linearly independent pure states  $\{   |\Psi_k\rangle\}_{k=1}^n$  and  a generic choice of prior probabilities~$\{  p_k\}_{k=1}^n$,  the success probability of the square root measurement can be expressed as \cite{pozza}
 \begin{align}
   P_{\rm SQ}   =    \sum_k  \,     \left(   \sqrt W \right)_{kk}^2 \, .  
 \end{align}  
The  convexity of the function $f(x)= x^2$ then implies the bound  $  P_{\rm SQ}  \ge \big(  \tr \sqrt W \, \big)^2/n $. 
 
\section{Greedy strategy and Bayesian updating}

Here we show that Bayesian updating gives the optimal greedy strategy introduced in the main text. This follows from the observation that  the optimal measurement (and the optimal guess) at step $s$ of the greedy strategy are determined solely by the posterior probability distribution after the measurement at step $s-1$, as will be explicitly shown at the end of this section.

To optimize the greedy strategy, we need to maximize ${\mathscr P}^{\rm G}_s=\sum_{r=1}^{n} \eta_{r}^{(s)} \langle   \Psi_{r} |E_s(r) |\Psi_r\rangle$ over all POVM measurements on particle~$s$, $\{E_s(r)\}_{r=1}^n$. 
Noticing that the source state $|\Psi_k\rangle$ restricted to particle~$s$ is $|\Psi_k\rangle_{\!s}=|0\rangle$ for $s< k$, and $|\Psi_k\rangle_{\!s}=|\phi\rangle$ for $s\ge k$, the following relations are self evident:  
%
\begin{eqnarray}
\label{ps-t}
{\mathscr P}^{\mathrm{G}}_{s}\!\!\!&=&\!\!\!
\sum_{r=1}^{s} \eta_{r}^{(s)}\,   \langle   \phi | E_s(r) |  \phi \rangle \! +\! \sum_{r=s+1}^n \eta_{r}^{(s)} \,   \langle   0 | E_s(r)|  0  \rangle  \nonumber\\
\!\!\!& \leq&\!\!p_{\phi}^{(s)} \!\sum_{r=1}^{s}  \langle   \phi | E_s(r) |  \phi  \rangle\!      +\! p_{0}^{(s)}\! \sum_{r=s+1}^n    \langle   0 | E_s(r)|  0  \rangle   \nonumber \\[.5em]
          \!\!\!&=&\!\! \! p_{\phi}^{(s)}\! \   \langle   \phi | \Pi_s(\phi)|  \phi  \rangle + p_{0}^{(s)}\,     \langle   0 | \Pi_s(0) |  0  \rangle ,
\end{eqnarray}
where 
$p_{\phi}^{(s)} :=\max_r{\{\eta_{r}^{(s)}\}_{r=1}^{s}}$, $p_{0}^{(s)}=\max_r{\{\eta_{r}^{(s)}\}_{r=s+1}^n}$,  $\Pi_{s}(\phi)=\sum_{r=1}^{s} E_s(r)$, 
and $\Pi_s(0)=\openone -\Pi_s(\phi)$.  
The inequality  is saturated by choosing a new POVM $\{E'_s(r)\}_{r=1}^n$ whose elements are non-zero only in the two positions that maximize the prior probabilities:
\begin{equation}
r_\phi=\argmax_{r}{\{\eta_{r}^{(s)}\}_{r=1}^{s}},\; r_0=\argmax_{r}{\{\eta_{r}^{(s)}\}_{k=s+1}^n},
\end{equation}
so that 
$E'_s ( r_0 )  =  \Pi_s (0)$   and $E'_s ( r_1 ) = \Pi_s (\phi ) $. This justifies the choice of priors in Eq.~(12) of the main text.

The success probability can now we written in terms of the Helstrom matrix $\Gamma_s=p_{\phi}^{(s)} |\phi\rangle\langle\phi|-p_{0}^{(s)} |0\rangle\langle0|$ as:
\begin{eqnarray}
{\mathscr P}^{\mathrm{G}}_{s}&=&p_{0}^{(s)}+ \tr\left(\Pi_{s}(\phi) \Gamma_s\right)\leq p_{0}^{(s)}+ \tr \left(\Gamma_s^{+}\right)\nonumber \\
&=&\frac{1}{2}\left( p_{\phi}^{(s)}+p_{0}^{(s)}+\tr|\Gamma_{s}|\right),
\end{eqnarray}
where $\Gamma_s^{+}$ is the positive part of matrix $\Gamma$. 
The inequality is saturated by choosing $\Pi_s(\phi)$ to be the projector onto the positive subspace of $\Gamma$~\cite{helstrom}.

We now show that the optimal measurement and guess at step $s$ of the greedy strategy 
do not depend on the particular sequence measurement outcomes, but only on the posterior probability distribution after the measurement at step $s-1$.
Let us  introduce the short-hand notation ${\mathbf{r}_{s}}:=\{r_{1},\ldots , r_{s}\}$ for a sequence of results obtained up to step~$s$.
The average success probability at each step $s$ is given by
 \begin{equation}
\label{ps-tbayes}
\sum_{k=1}^n\sum_{\mathbf{r}_{s}}p(\mathbf{r}_{s},k) \delta_{k\,\hat{k}(\mathbf{r}_{s})}
\leq \sum_{\mathbf{r}_{s}}\max_{k}{p(\mathbf{r}_{s},k)},
\end{equation}
where $p(\mathbf{r}_{s},k)$ is the joint probability of obtaining the sequence $\mathbf{r}_{s}$ of results and the change point occuring at position $k$, and $\hat{k}(\mathbf{r}_{s})\in \{1,\ldots, n\}$ is the decision function that assigns to each $\mathbf{r}_{s}$ the guessed change point position~$k=\hat{k}(\mathbf{r}_{s})$. The inequality can be saturated by
$\hat{k}(\mathbf{r}_{s})=\argmax_{k} p(\mathbf{r}_{s},k)$. Since the source states $|\Psi_k\rangle$   are of product form, we can write 
\begin{equation}
p(\mathbf{r}_{s},k)={1\over n}p(\mathbf{r}_{s-1}| k)\,\langle\Psi_k|E_{s}(r_s) |\Psi_k\rangle ,
\end{equation}
%
where we recall that  $|\Psi_k\rangle$ restricted to particle $s$ is $|\Psi_k\rangle_{\!s}=|0\rangle$ for $s< k$, and $|\Psi_k\rangle_{\!s}=|\phi\rangle$ for $s\ge k$.  The measurement over the $s$-th particle is represented by
the POVM $\{E_{s}(r)\}_{r=1}^n$ and  it is understood that it may depend on the  sequence~$\mathbf{r}_{s-1}$ of previous results.  Hence,  
the optimal greedy average success probability at step~$s$, can be written as
\begin{equation}
P^{\rm G}_{s}= 
\sum_{\mathbf{r}_{s-1}}p(\mathbf{r}_{s-1}){\mathscr P}^{\rm G}_{s}({\mathbf{r}_{s-1}}),
\end{equation}
where the probability of successful identification of the change point at step $s$ conditioned to the occurrence of the sequence $\mathbf{r}_{s-1}$ [${\mathscr P}^{\rm G}_s$ in 
Eq.~(\ref{ps-t}); we recall that the dependence on ${\bf r}_{s-1}$ is understood there]~is 
\vspace{-.2cm}
\begin{eqnarray}
{\mathscr P}^{\rm G}_{s}({\mathbf{r}_{s-1}})\!\!&=&\!\!\max_{\{E_{s}(r)\}}\sum_{r=1}^n\max_{k} p(k|\mathbf{r}_{s-1})\nonumber
\\
\!\!&\times&\!\!\langle\Psi_k|E_{s}(r) |\Psi_k\rangle,
\label{only priors}
\\[-.5em]
\nonumber
\end{eqnarray}
\\[-1em]
and we have used Bayes' rule to  obtain the relation $(1/n)\,p(\mathbf{r}_{s-1}|k)=p(k|\mathbf{r}_{s-1}) p(\mathbf{r}_{s-1})$. From Eq.~(\ref{only priors}), it is apparent that  the optimal measurement can only depend on the updated priors $\eta_{k}^{(s)}:=p(k|\mathbf{r}_{s-1})$, rather than on the whole sequence of previous results, as the maximization is only subject to the POVM conditions $E_s(r)\ge0$ and $\sum_{r=1}^n E_s(r)=\openone$.  Likewise, the optimal guess can only depend on $\eta_{r}^{(s)}$ [Eq.~(\ref{ps-tbayes}) and the paragraph below~it].

\bibliographystyle{unsrt}

\end{document}